%% file: conf_ul_csi_isac.tex
\documentclass[10pt, conference]{IEEEtran}
\IEEEoverridecommandlockouts

\usepackage{graphicx} 
\graphicspath{{./figs/}}

\usepackage{subcaption}
\usepackage{amsmath, amssymb, amsfonts}
\allowdisplaybreaks
\usepackage{xcolor}
\usepackage{float}

\usepackage{url} 

\usepackage[acronym]{glossaries} 
\loadglsentries[acronym]{acronyms}

\newcommand\copyrighttext{%
  \textcopyright \the\year{} IEEE.
  Personal use of this material is permitted. \\
  Permission from IEEE must be obtained for all other uses, including reprinting/republishing this material for advertising or promotional purposes, collecting new collected works for resale or redistribution to servers or lists, or reuse of any copyrighted component of this work in other works.
}

\begin{document}

\bstctlcite{BSTcontrol}

\title{Evolution Toward Sensing-Enabled RAN: \\ Practical Bistatic Sensing via Uplink}

\author{
    \IEEEauthorblockN{
        Kenan Turbic\IEEEauthorrefmark{1},
        Khanh-Toan Pham\IEEEauthorrefmark{1},
        S\l{}awomir Sta\'nczak\IEEEauthorrefmark{1}\IEEEauthorrefmark{2}
        }%
	\IEEEauthorblockA{\IEEEauthorrefmark{1}
            Fraunhofer Heinrich Hertz Institute (HHI),
            Berlin, Germany \\
        }%
        \IEEEauthorblockA{\IEEEauthorrefmark{2}
            Technische Universit{\"a}t Berlin, Germany
        }%
        \IEEEauthorblockA{
            kenan.turbic@hhi.fraunhofer.de
        }%
        \thanks{This work was partially supported by the Federal Ministry of Research, Technology and Space (BMFTR, Germany) in the “Souverän. Digital. Vernetzt.” programme, joint projects xG-RIC (grant numbers: 16KIS2429K and 16KIS2434) and SENSATION (grant number: 16KIS2526).}
}%

\IEEEpubid{%
  \begin{minipage}{0.8\textwidth}\ \\[10mm] \centering
      \footnotesize \copyrighttext
  \end{minipage}
}

\maketitle

\begin{abstract}
This paper considers uplink \gls{csi}-based sensing as an opportunistic approach to enabling sensing in cellular networks, with minimum required modifications to the infrastructure.
A practical sensing signal processing procedure is presented, combining \acrshort{5g} positioning function with multi-antenna array processing for passive target localization.
The approach is validated through simulation and practical experiments with a \acrlong{sdr}-based setup, utilizing open-source software frameworks and commercially available hardware components.
Results demonstrate feasibility of the uplink \gls{csi}-based sensing for detection, localization and tracking of moving targets over local areas.
\end{abstract}


\glsresetall 


\section{Introduction}
\label{sec:intro}
The ubiquitous and dense deployment of cellular communication networks creates an infrastructure of substantial scale.
Exploiting this infrastructure for sensing, in addition to its primary communication function, is the fundamental driver for implementing the \gls{isac} concept into the next generation wireless networks \cite{Prelcic2024}.

While research into new waveforms and frequency bands is important for the long-term vision toward achieving the full \gls{isac} potential \cite{LiuF2026}, an evolutionary path toward sensing-enabled networks is a more realistic near-term route to practical deployment.
Such an approach requires minimum modifications to the \gls{ran}, thereby reducing the barrier to adoption.

A particularly resource-efficient realization of this evolutionary approach is to reuse the known pilot signals transmitted over the communication links.
These signals are used for channel estimation needed for data equalization and user scheduling \cite{Dahlman2021book}, with the obtained \gls{csi} being discarded afterwards.
Processing these estimates additionally for sensing, rather than discarding them, incurs almost no additional cost and requires practically no changes to the \gls{ran} standard.

Bistatic sensing with \glsunset{5g}\gls{5g} signals has received growing attention recently \cite{Giroto2026pp}.
Most reported studies are simulation-based, while exploring sensing opportunities in both downlink \cite{Khosroshahi2024, Nataraja2025} and uplink (\glsunset{gnb}\gls{gnb}) \cite{Tapio2024}. 
The few experimental studies typically depart from the \gls{5g} standard, using generic \gls{ofdm} waveform \cite{Brunner2025}, WiFi signals \cite{Sanson2026, Pegoraro2024}, or  non-standard pilots \cite{Kanth2026}.

Real-time standard-compliant demonstrations of \gls{csi}-based \gls{isac} remain scarce.
Recently, NVIDIA Aerial \gls{sdr} \gls{gnb} implementation has been used for real-time demonstration of \gls{ue} positioning \cite{Wiesmayr2025} and passive target sensing \cite{Villa2026pp} via uplink \gls{csi} based on \gls{dmrs}, using a trained neural network model for moving person localization.

In this work, we consider uplink \gls{csi}-based bistatic sensing in evolved 5G systems and present a simple and practical approach to enable target localization.
We develop a complete signal processing procedure, combining \gls{5g} \gls{ue} positioning with multi-antenna array processing.
The approach is validated through simulation and preliminary measurements, obtained with a real-time \gls{isac} testbed based on NVIDIA Aerial, \glsunset{3gpp}\gls{3gpp}-compliant Open \glsunset{ran}\gls{ran} \gls{gnb} implementation in \gls{sdr}.

The rest of this paper is organized as follows.
The uplink \gls{csi}-based sensing is discussed in Section~\ref{sec:bistatic}, while the target localization method is developed in Section~\ref{sec:5g_sensing}.
Simulation-based validation is presented in Section~\ref{sec:sim} and preliminary experimental results are provided in Section~\ref{sec:exp}.
The concluding remarks are provided in Section~\ref{sec:conclusions}.

\IEEEpubidadjcol

\section{Uplink CSI-Based Sensing} 
\label{sec:bistatic}

\subsection{Channel Model}
\label{sec:bistatic_model}
In \gls{5g} uplink transmissions, known pilot signals are embedded within the transmitted \gls{ofdm} waveform, to enable \gls{csi} estimation at the \gls{gnb}. 
Rather than discarding the \gls{csi} after serving its communication purpose, it can be further processed to extract information about the propagation environment.

Two types of \gls{5g} uplink pilot signals are suitable for sensing, i.e., \gls{dmrs} and \gls{srs} \cite{3GPP38211}.
The \gls{dmrs} is embedded within every uplink data transmission and the obtained \gls{csi} is used for channel equalization prior to symbol detection, while the \gls{srs} is transmitted independently of user data and is used to support scheduling decisions.

For single-antenna \gls{tx} and \gls{rx}, the \gls{ctf} observed at the \gls{gnb} can be expressed as \cite{Sturm2011} 
\begin{align}
    h(f_n, t_m)
    = \sum_{l=0}^{N_p-1} \beta_l  \, e^{-j 2 \pi f_n \tau_l} \, e^{j 2 \pi f_{D}^l t_m}
    \label{eq:ctf_siso}
\end{align}
where
\begin{description}
    \item[$N_p$] number of multipath components arriving at the \gls{rx};
    \item[$t_m$] time instances of pilot-bearing \gls{ofdm} symbols, \\ i.e., $t_m = m \, T_p$;
    \item[$T_p$] pilot symbol periodicity;
    \item[$f_n$] $n$-th subcarrier frequency, i.e., $f_n = f_0 + n \Delta f$;
    \item[$f_0$] carrier frequency;
    \item[$\Delta f$] subcarrier spacing;
    \item[$\beta_l$] complex path gains;
    \item[$\tau_l$] propagation delays;
    \item[$f_{D}^{l}$] exhibited Doppler shifts.
\end{description}

The channel includes contributions from multiple propagation paths between the \gls{tx} and the \gls{rx}, including the \gls{los} (if unobstructed) and multipath components associated with objects in the environment, i.e., both clutter and targets of interest.
The complex gains account for the propagation losses and phase shifts exhibited upon scattering, the delays are associated with propagation path lengths, while Doppler shifts result from scatterer and/or \gls{ue} motion.
In this work we assume that the \gls{ue} is nomadic, i.e., stationary during the observation period, and the Doppler shifts are solely due to scatterer motion.
However, the implications of \gls{ue} mobility are discussed in Section~\ref{sec:challenges}.

\begin{figure}[!t]
	\centering
    \includegraphics[width=0.98\columnwidth]{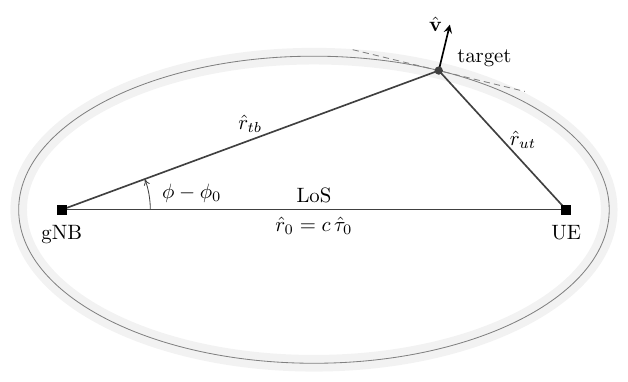}
	\caption{Uplink channel propagation geometry.}
	\label{fig:geom_bistatic}
\end{figure}

The sensing context follows from the uplink propagation geometry, illustrated in Fig.~\ref{fig:geom_bistatic}.
A simple scenario is considered for simplicity, including the \gls{gnb}, \gls{ue} and a single scatterer assumed to be the target of interest.
As shown in the figure, for a given echo signal delay, the corresponding target lies on an ellipse with the \gls{ue} and \gls{gnb} at its foci, and the semi-major axis being twice the propagation path length \cite{Parsons2000book}.
The Doppler shift corresponds to the rate at which the total propagation path length changes over time and is proportional to the relative speed in the direction normal to the propagation ellipse \cite{Willis2005book}.

As observed from \eqref{eq:ctf_siso}, propagation delays and Doppler shifts yield (independent) harmonic components across the frequency and time samples, respectively, and thereby can be estimated via Fourier analysis of the \gls{ctf} in the two dimensions \cite{Sturm2011}.
An \gls{idft} over the frequency dimension yields a sequence of \gls{cir} snapshots, while a successive (direct) \glsunset{dft}\gls{dft} over the time dimension produces a delay-Doppler map.
However, limited interpretation of this map in terms of absolute spatial reference is possible from the \gls{csi} alone. 

\subsection{Sensing Limitations and Opportunities}
\label{sec:bistatic_opplim}
It is important to point out that the \gls{rx} performs time synchronization relative to the first arriving signal component, i.e., the \gls{los} path, if unobstructed.
As a result, the path delays observed in the estimated \gls{cir} are relative to this synchronization reference, rather than to the (unknown) absolute transmission time.
Combined with the generally unknown position of the mobile \gls{ue}, this means that the absolute bistatic range to a target cannot be established directly from the \gls{csi}.
Consequently, target position estimation is not possible without additional information.
 
Despite this limitation, useful sensing capabilities remain.
The \gls{cir} variations caused by moving objects in the environment, can be detected without any knowledge of absolute distances or \gls{ue} position.
This enables moving target presence detection, a capability of practical interest for use cases such as intrusion detection in private or campus network deployments.
It can be implemented with minimal additional cost, without modifications to the \gls{ran} standard or the hardware.

\section{Enabling Target Localization}
\label{sec:5g_sensing}
 
\subsection{Total Bistatic Range Recovery}
\label{sec:range}
As  previously, the path delays in the \gls{csi} estimated at the \gls{gnb} are relative to its established synchronization reference.
However, if the absolute delay of the \gls{los} path is known, the physical delay of any path can be recovered as
\begin{align}
    \hat{\tau} = (\hat{\tau}_0 - \tau_0) + \tau
    \label{eq:phys_delay}
\end{align}
where
\begin{description}
    \item[$\hat{\tau}_0$] estimated physical \gls{los} delay;
    \item[$\tau_0$] \gls{los} delay in the observed \gls{cir};
    \item[$\tau$] target echo delay in the observed \gls{cir}.
\end{description}

The physical \gls{los} delay follows directly in scenarios when the \gls{ue} position is known.
This could be the case if a reference \gls{ue} is specifically deployed for sensing, or if a fixed-position \gls{cpe} is used.
Alternatively, two \gls{5g} mechanisms present themselves as candidates for absolute \gls{los} delay estimation, i.e., \gls{ta} and \gls{5g} \gls{ue} positioning \cite{Dahlman2021book}.

\gls{ta} is a \gls{mac}-layer mechanism by which the \gls{gnb} estimates the round-trip propagation delay to the \gls{ue} and issues timing adjustment commands to align uplink transmissions with the reception window \cite{3GPP38321}.
However, the accuracy of the \gls{ta} is limited by its coarse quantization and the narrow bandwidth of the utilized \gls{prach} signal, resulting in a distance error on the order of tens of meters, hence not suitable for target localization.
The \Gls{5g} uplink-based \gls{ue} positioning function is a more accurate alternative for estimating the physical \gls{los} path length \cite{ChaHS2025}.
It relies on the \gls{srs}, which can span the entire uplink bandwidth and therefore provides a considerably finer time resolution than the \gls{prach} signal.

\subsection{Angle-of-Arrival Estimation}
\label{sec:aoa}
The bistatic target path length alone is insufficient for target localization. 
This is because the corresponding propagation ellipse (Fig.~\ref{fig:geom_bistatic}), can be arbitrarily oriented in space and the scatterer position on this ellipse remains unknown.
Target localization additionally requires \gls{aoa} estimation at the \gls{gnb}, by exploiting multiple \gls{rx} antennas.
 
The \gls{los} path \gls{aoa} establishes the orientation of the propagation ellipse in space relative to the known \gls{gnb} array broadside direction, while the target echo \gls{aoa} identifies the position of the scatterer on the ellipse.
Jointly, \gls{aoa} estimates together with the \gls{los} range enable target localization.

For multiple \gls{rx} antennas at the \gls{gnb}, the \gls{ctf} across the array elements can be expressed as%
\footnote{Here we adopt the typical assumption that the target speed is low enough such that the fast-time Doppler shift is negligible.}
\begin{align}
        \mathbf{h}(f_n, t_m)
        &= \sum_{l=0}^{N_p-1} \beta_l  \, e^{-j 2 \pi f_n \tau_l} \, e^{j 2 \pi f_{D}^l t_m} \, \mathbf{a}(\phi_l)
        \label{eq:ctf_simo}
\end{align}
where
\begin{description}
    \item[$\phi_l$] path azimuth \glspl{aoa};
    \item[$\mathbf{a}$] array steering vector.
\end{description}
For an \gls{ula}, the steering vector is
\begin{align}
    \mathbf{a}(\phi)
    &= \left[ 1, e^{-j 2 \pi \frac{d}{\lambda} \sin\phi}, \ldots, e^{-j 2 \pi (N_a - 1) \frac{d}{\lambda} \sin\phi} \right]^T
\end{align}
where
\begin{description}
    \item[$N_a$] number of antenna elements;
    \item[$d$] inter-element spacing, i.e., typically $d = \lambda / 2$;
    \item[$\lambda$] carrier wavelength.
\end{description}

The \gls{ctf} at the $n$-th sub-carrier, $m$-th pilot OFDM symbol, and $k$-th antenna element can be written as
\begin{align}
        h[n, m, k]
        &= \sum_{l=0}^{N_p-1} \beta_l  \, e^{-j 2 \pi f_0 \tau_l} \, e^{-j 2 \pi \tau_l \Delta f n} \, \times \notag \\
        &\hspace{2cm} e^{j 2 \pi f_{D}^l T_p m} \, e^{-j 2 \pi \frac{d}{\lambda} \sin\phi_l \, k}
        \label{eq:ctf_simo_discrete}
\end{align}

As follows, in addition to the harmonic terms in frequency and time dimensions, the arriving signals also result in a harmonic component across the antenna index, with frequencies determined by the respective \glspl{aoa}.
Therefore, an additional \gls{idft} across the antenna dimension can be applied to the \gls{ctf} to estimate the \glspl{aoa} \cite[Ch.~9]{Richards2022book}.
In addition to the simple \gls{dft}-based approach, superior \gls{aoa} estimation methods can be employed, e.g., Capon beamformer \cite{Capon1969}, \acrshort{music} \cite{Schmidt1986} etc., at the expense of higher computational complexity.

\subsection{Target Position Mapping}
\label{sec:position}
With the absolute target echo delay and \gls{aoa} estimated, and the absolute \gls{los} delay and \gls{aoa} used to establish the propagation ellipse geometry, the target position is obtained at the intersection of the \gls{aoa} ray from the \gls{gnb} and the propagation ellipse.
The target distance from the \gls{gnb} is then obtained as \cite[Ch.~3]{Willis2005book}
\begin{align}
    \hat{r}_{tb}(\hat{\tau}, \hat{\phi}; \, \hat{\tau}_0, \hat{\phi}_0)
    &= \frac{a \, (1 - \zeta^2)}{1 - \zeta \, \cos\Delta\phi}
    \label{eq:target_dist}
\end{align}
where
\begin{description}
    \item[$\Delta\phi$] target \gls{aoa} relative to the \gls{los}, i.e., $\Delta\phi = \phi - \phi_0$;
    \item[$\hat{\phi}$] target \gls{aoa} estimate;
    \item[$\hat{\phi}_0$] \gls{los} component \gls{aoa} estimate;
    \item[$a$] major semi-axis of the ellipse, i.e., $a = c\,\hat{\tau}/2$;
    \item[$\zeta$] eccentricity of the ellipse, i.e., $\zeta = c\,\hat{\tau}_0/2$;
    \item[$c$] speed of light.
\end{description}
This distance and the target echo \gls{aoa} yield the target position in Cartesian coordinates through polar to Cartesian transformation, assuming known \gls{gnb} position and array orientation.

The established propagation ellipse geometry also determines the direction of the target velocity component estimated from the bistatic Doppler analysis.
In particular, this direction corresponds to the ellipse normal at the target location, with the unit vector given as
\begin{align}
    \hat{\mathbf{v}}(\hat{\phi}; \, \hat{\tau}_0, \hat{\phi}_0)
    &= \frac{1}{\sqrt{1 + 2\zeta\cos\Delta\phi + \zeta^2}}
    \begin{bmatrix}
        \cos\Delta\phi - \zeta \\
        \sin\Delta\phi
    \end{bmatrix}
\end{align}

\begin{figure*}[ht!]
	\centering
	\includegraphics[scale=1]{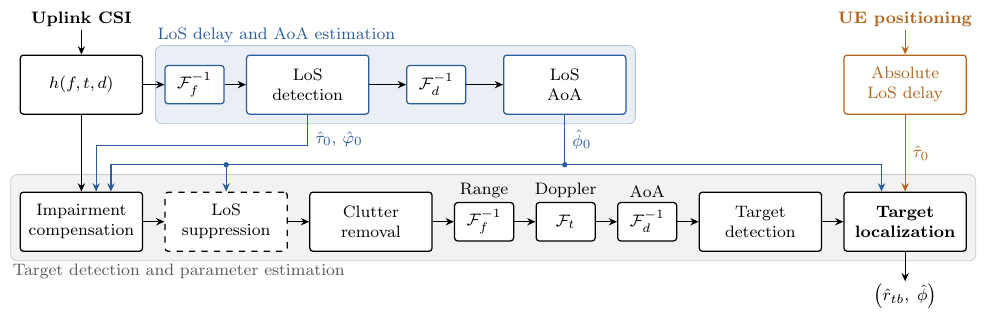}
	\caption{Uplink \gls{csi} signal processing pipeline at the \gls{gnb}.}
	\label{fig:rx_procedure}
\end{figure*}

\subsection{Sensing Procedure Outline}
\label{sec:procedure}
The \gls{gnb}-side sensing signal processing procedure is shown in Fig.~\ref{fig:rx_procedure}, where the \gls{ctf} from the uplink \gls{csi} is processed in two branches.
The first one estimates \gls{los} parameters to support synchronization, hardware impairments compensation, \gls{los} suppression and target localization.
The second one performs \gls{los} and clutter suppression, and Fourier transform steps to obtain the range-Doppler-angle map for target detection and localization, utilizing inputs from the first branch and \gls{ue} positioning mechanisms.

In the \gls{los} processing pipeline, the \gls{cir} is first obtained via an \gls{idft} across the frequency axis for a single antenna element and a single \gls{ofdm} pilot symbol.
The \gls{los} delay bin is identified as the least-delay peak and an \gls{idft} is performed across the signals from all antenna elements in the given bin, with the subsequent peak detection yielding the \gls{los} \gls{aoa}.

Once available, the \gls{los} \gls{aoa} estimate can be additionally used to suppress the strong \gls{los} contribution from the vector \gls{ctf} estimate.
This can be achieved via null-subspace projection method \cite{Friedlander2012}, i.e., by applying the following operation
\begin{align}
    \tilde{\mathbf{h}}(f_n, t_m) = \left( \mathbf{I} - \frac{1}{N_a} \mathbf{a}(\hat{\phi}_0) \, \mathbf{a}(\hat{\phi}_0)^H \right) \, \mathbf{h}(f_n, t_m)
    \label{eq:los_suppress}
\end{align}
where
\begin{description}
    \item[$\hat{\phi}_0$] \gls{los} \gls{aoa} estimate;
    \item[$\mathbf{I}$] identity matrix, i.e., $N_a \times N_a$;
    \item[$(\,.\,)^H$] complex-conjugate (Hermitian) transpose.
\end{description}
The same approach can be applied to suppress interference from external sources, if their \glspl{aoa} are properly estimated.

For moving target detection, the static clutter is then suppressed by subtracting the \gls{ctf} time-average, i.e., 
\begin{align}
    \tilde{\mathbf{h}}^\dagger(f_n, t_m) =  \tilde{\mathbf{h}}(f_n, t_m) - \frac{1}{N_s} \sum_{m=1}^{N_s} \tilde{\mathbf{h}}(f_n, t_m)
\end{align}
where
\begin{description}
    \item[$N_s$] number of pilot OFDM symbols in the \gls{cpi}.
\end{description}
This effectively removes the zero-Doppler component, thereby facilitating the subsequent processing and target detection.

Following the \gls{los} and clutter suppression steps, the range-Doppler-angle processing steps are applied to the \gls{ctf} estimates, involving successive Fourier transforms%
\footnote{The continuous-variable expressions are provided for simplicity, while discrete versions are actually implemented.}
\begin{align}
    h_1(\tau, t, d) &= \mathcal{F}_f^{-1} \left\{ \tilde{h}^\dagger(f, t, d) \right\}
    \label{eq:cir_seq}
    \\
    h_2(\tau, f_D, d) &= \mathcal{F}_t \left\{ h_1(\tau, t, d) \right\}
    \label{eq:dd_map}
    \\
    h_3(\tau, f_D, d) &= \mathcal{F}_{d}^{-1} \left\{ h_2(\tau, f_D, d) \right\}
    \label{eq:dda_map}
\end{align}
where $d$ is the spatial dimension, i.e., distance along the array aperture relative to the reference point (e.g., first element).

The delay-Doppler-angle map \eqref{eq:dda_map} is processed for target detection and parameter estimation by employing detection methods, e.g., \gls{cfar} \cite[Sec.~6.5]{Richards2022book}, followed by subsequent tracking and data association steps to estimate target trajectories across multiple \glspl{cpi} (not shown).
To reduce the impact of the straddle loss, the Fourier transforms \eqref{eq:cir_seq}\,-\,\eqref{eq:dda_map} can be implemented with zero-padding to achieve interpolation in the respective domains.
Alternatively, a more computationally-efficient second-order polynomial interpolation can be applied instead \cite[Sec.~5.4.6]{Richards2022book}.

Finally, by employing the absolute \gls{los} delay estimate obtained via the \gls{5g} \gls{ue} positioning function, the relative target echo delays are translated into the target distances from the \gls{gnb}, according to \eqref{eq:target_dist}, thereby completing the target localization procedure.

\section{Practical Challenges}
\label{sec:challenges}
Several practical challenges are associated with the presented uplink sensing approach.
These are briefly discussed, with an indication of possible mitigation approaches.

\subsubsection{Limited Tx power}
The sensing range of the uplink approach is limited by the low \gls{ue} transmit power and by the strong direct path, which masks weak target echoes \cite{Tapio2024}.
Accordingly, this approach is better suited for short-range scenarios, e.g., indoor and industrial environments.
The range can be extended with \gls{cpe} devices, which support higher power classes and higher-gain antennas than handheld \glspl{ue}.

\subsubsection{Positioning error}
The accuracy of the target position estimate is jointly determined by the errors in all estimates entering the expression \eqref{eq:target_dist}.
The absolute \gls{los} delay estimation is governed by the accuracy of the \gls{5g} \gls{ue} positioning, constrained by the system bandwidth, \gls{snr}, channel conditions, and the number of \glspl{gnb} involved.
The errors in the \gls{aoa} estimates for both the \gls{los} path and the target echo depend on the array size (aperture), the \gls{snr}, and the adopted estimation method.

\subsubsection{UE mobility}
The presented method assumes that the \gls{ue} is nomadic and remains static over the \gls{cpi}, such that the \gls{los} component and clutter can be treated as time-invariant.
In general, the impact of \gls{ue} motion must be compensated for, i.e., by estimating the \gls{los} path Doppler shift and employing methods typically used in automotive radar \cite{Richards2022book}.

\subsubsection{CSI availability and signaling overhead}
The \gls{dmrs}-based estimates are available whenever data is being transmitted on the uplink, and thus can be sparse or irregular under low traffic load conditions.
In contrast, \gls{srs} is scheduled independently of data transmissions and offers more flexible time-frequency configuration, at the expense of occupying additional uplink resources.
A combined opportunistic strategy could be of interest, where \gls{dmrs}-based \gls{csi} estimates are exploited in high traffic conditions, while \gls{srs} transmissions are scheduled during periods of low data activity.

\subsubsection{Sensing with communication hardware}
The hardware impairments, including \acrlong{to}, \acrlong{cfo}, and \gls{sfo} between the \gls{tx} and the \gls{rx}, limit the coherence of the \gls{rx} signal over time \cite{Giroto2026pp}.
While these impairments have a negligible impact over the short symbol blocks processed for communication data detection, sensing low speed targets (e.g., pedestrians) requires extended \glspl{cpi}, spanning multiple radio frames, over which even small impairments accumulate.
However, presence of the \gls{los} or another known strong path can be exploited to track and compensate for this drift \cite{Pegoraro2024, Giroto2025}.

\section{Simulation Results}
\label{sec:sim}
For initial validation of the approach, we simulate a scenario consisting of a single-antenna \gls{ue} communicating with a \gls{gnb}, equipped with a four-element \gls{ula} with half-wavelength element spacing.
The \gls{ue} is placed at a distance of 10\,m from the \gls{gnb}, in the direction of the array broadside ($\phi_0 = 0^\circ$), and is assumed to be static over the sensing interval.
A single passive target is considered at a distance of 15\,m and an azimuth angle of $-30^\circ$ relative to the array broadside.
The target is assumed static for simplicity.

The signal propagation is assumed to be dominated by the \gls{los} path and the single-scatter target echo, with the \gls{ctf} at each antenna element simulated according to~\eqref{eq:ctf_simo_discrete}, with the (relative) delays and \glspl{aoa} determined by the scenario geometry.
The \gls{los} path \gls{snr} at the \gls{rx} antenna output (prior to signal processing gain) is set to 20\,dB, with the target echo being 12\,dB weaker.

The \gls{5g} system operates in the \glsunset{fr1}\gls{fr1} band at a carrier frequency of 3.75\,GHz, with 4096 \gls{ofdm} subcarriers and a subcarrier spacing of 30\,kHz.
The \gls{rx} synchronization reference is arbitrarily assumed to precede the \gls{los} path by 2 sampling intervals.
Error-free \gls{ue} position estimation is assumed, the exact absolute \gls{los} delay being used for target-\gls{gnb} distance estimation according to \eqref{eq:target_dist}.

The signal processing follows the procedure outlined in Section~\ref{sec:procedure}.
The \gls{ofdm} system, scenario and signal processing parameters are summarized in Table~\ref{tab:ofdm_sim_params}.
\begin{table}[!t]
    \centering
    \caption{System and scenario parameters in the simulation.}
    \label{tab:ofdm_sim_params}
    \renewcommand{\arraystretch}{1.2}
    \begin{tabular}{|c|c|c|}
        \hline
        \textbf{Parameter} & \textbf{Symbol} & \textbf{Value} \\
        \hline\hline
        Num. subcarriers & $N_{sc}$ & 4096 \\ \hline
        Num. Rx antennas & $N_a$ & 4 \\ \hline
        Subcarrier spacing & $\Delta f$ & 30\,kHz \\ \hline
        Carrier frequency & $f_0$ & 3.75\,GHz \\
        \hline\hline
        UE distance & $r_0$ & 10\,m \\ \hline
        UE AoA & $\phi_0$ & $0^\circ$ \\ \hline
        LoS SNR & $\rho_{0}$ & 20\,dB \\
        \hline\hline
        Target distance & $r_{tb}$ & 15\,m \\ \hline
        Target AoA & $\phi$ & $-30^\circ$ \\ \hline
        Target SNR & $\rho$ & 8\,dB \\
        \hline\hline
        Range interpolation factor & $N_{int}^{\tau}$ & 6 \\ \hline
        AoA interpolation factor & $N_{int}^{\phi}$ & 6 \\
        \hline
    \end{tabular}
\end{table}
 
\begin{figure}[!t]
	\centering
	\includegraphics[width=0.95\columnwidth]{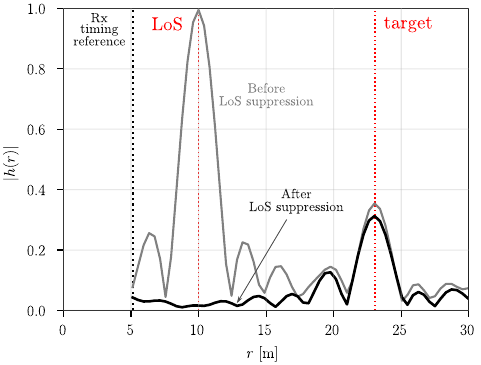}
	\caption{Channel impulse response at the \gls{rx} (range-corrected).}
	\label{fig:cir_observed}
\end{figure}

Fig.~\ref{fig:cir_observed} shows the \gls{cir} at one of the \gls{gnb} antennas, before and after the \gls{los} suppression with range correction.
By utilizing the knowledge of the absolute \gls{los} distance, the range axis in the figure is adjusted to show the total bistatic propagation path length.
The \gls{cir} curves start at the \gls{rx} synchronization reference, i.e., the minimum observable range%
\footnote{All targets along the \gls{los} line will appear at this range, and cannot be range-resolved in a bistatic setting \cite{Willis2005book}.}. 
The \gls{los} path is observed as the strongest component at 10\,m, while the target echo appears as a weaker component with path length of around 23\,m.
The \gls{los} suppression step effectively removes the strong \gls{los} component, thereby facilitating the subsequent processing and target detection.
\begin{figure}[!t]
    \centering
    \subfloat[]{
        \includegraphics[width=0.48\columnwidth]{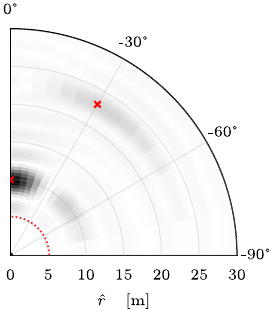}
        \label{fig:rng_ang_map_a}
    }
    \subfloat[]{
        \includegraphics[width=0.48\columnwidth]{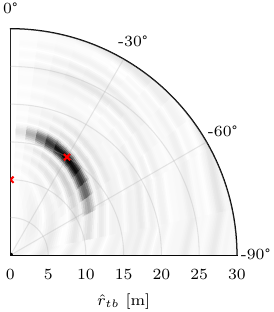}
        \label{fig:rng_ang_map_b}
    }
    \caption{Simulated range-angle map a) before and b) after \gls{los} suppression and monostatic range conversion.}
    \label{fig:rng_ang_map}
\end{figure}

Fig.~\ref{fig:rng_ang_map} shows the range-angle maps%
\footnote{The colormap is normalized to a linear intensity gradient from 0 (white) and the maximum value (black), separately in each plot.}
before (Fig.~\ref{fig:rng_ang_map_a}) and after (Fig.~\ref{fig:rng_ang_map_b}) \gls{los} suppression and bistatic range conversion to distance from the \gls{gnb}.
Fig.~\ref{fig:rng_ang_map_a} shows the total bistatic range (corrected), where the red circle indicates the minimum observable range, corresponding to the \gls{rx} synchronization reference (i.e., indicated by a black dotted line in Fig.~\ref{fig:cir_observed}), while the red crosses in the figure indicate true values for the \gls{los} and target echo paths.
As we observe, the corresponding peaks in the range-angle map appear correctly centered.

The impact of the \gls{los} suppression is clearly visible in Fig.~\ref{fig:rng_ang_map_b}, where the peak corresponding to the \gls{los} path is effectively removed and only the target appears in the map, at the exact target position specified as the simulation input.
The shape of the peak is slightly distorted, due to the bistatic to monostatic (relative to the \gls{gnb}) range conversion.
It should be pointed out that the whole observed space transformation is performed here for visualization purposes only, while the bistatic range correction and transformation to gNB-centric coordinates is practically performed only for detected targets.

\section{Experimental Results}
\label{sec:exp}
\begin{figure}[!t]
    \centering
    \subfloat[]{
        \includegraphics[height=4.9cm]{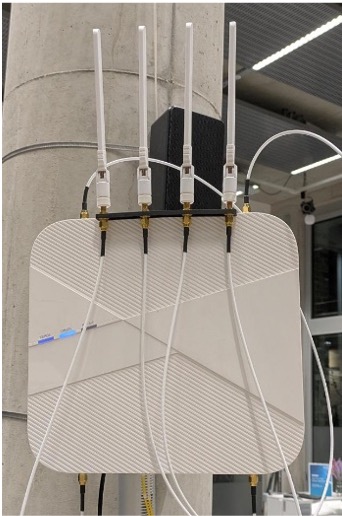}
        \label{fig:measurement_setup}
    }
    \subfloat[]{
        \includegraphics[height=4.9cm]{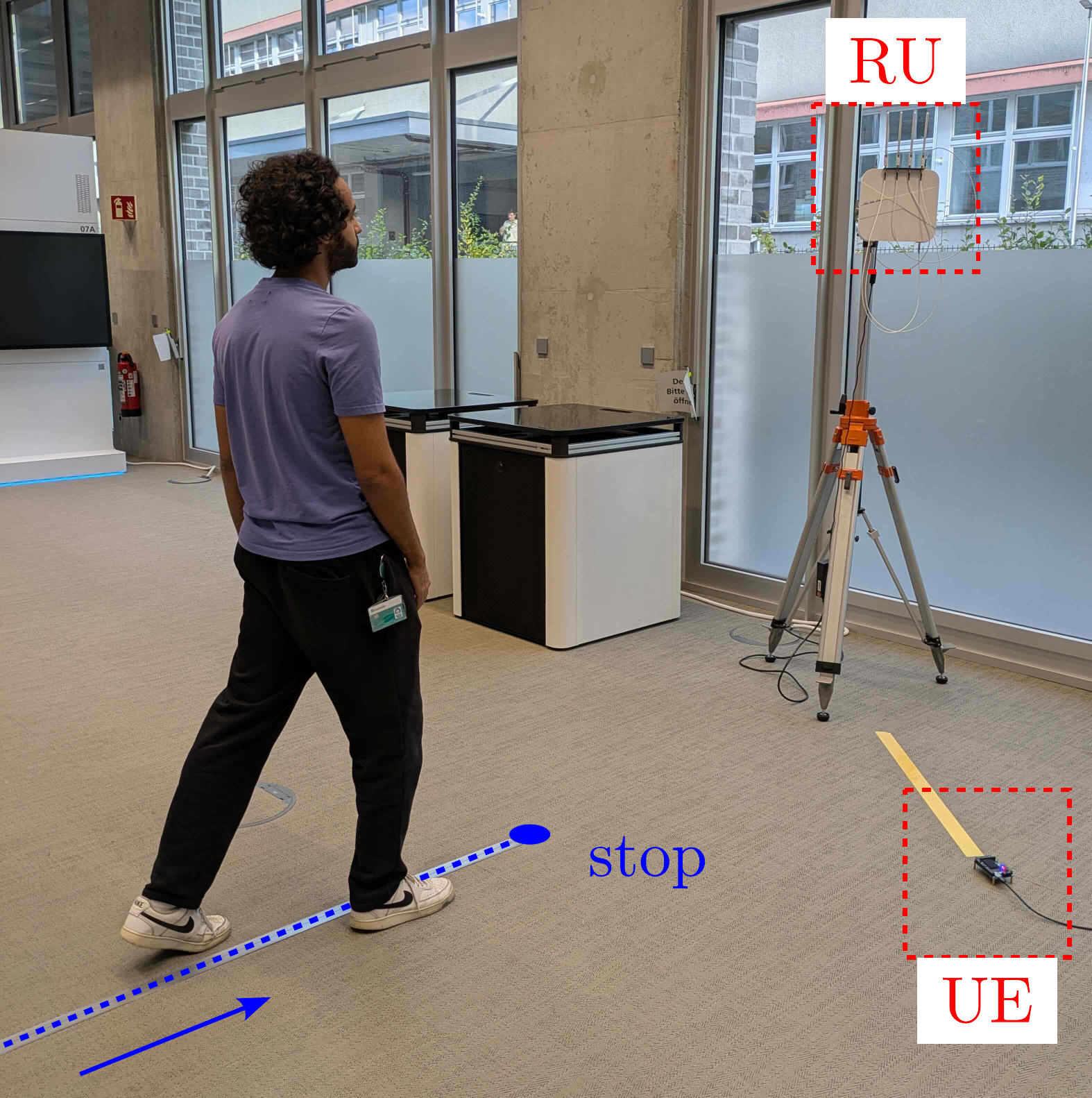}
        \label{fig:measurement_scenario}
    }
    \caption{Experimental setup a) \gls{ru} with a 4-element \gls{ula} and b) measurement scenario.}
    \label{fig:measurement}
\end{figure}

The presented sensing approach is further validated based on measurements with the NVIDIA Aerial \gls{sdr} \gls{gnb} implementation \cite{nvidia_aerial}.
The radio front-end in the experimental setup consists of a Foxconn RPQN \gls{ru} with 4 external antennas, arranged into a horizontal \gls{ula} with half-wavelength spacing by using a custom 3D-printed alignment structure, as shown in Fig.~\ref{fig:measurement_setup}.
The setup operates at a carrier frequency of 3.75\,GHz, with 3276 active subcarriers with a 30\,kHz spacing, occupying a total channel bandwidth of 100\,MHz. 

A Quectel RM500Q-GL \gls{5g} modem is used as the \gls{ue}, placed at a 2\,m distance from the \gls{ru} in the array broadside direction.
It is transmitting with maximum \gls{tx} power of 23\,dBm.
The \gls{ru} is set at a height of 2\,m, while the \gls{ue} is placed on the floor.
Acting as a sensing target, a subject was walking on a straight line towards the \gls{ru}, over distances from 6\,m to 3\,m, at $-32^\circ$ azimuth direction (Fig.~\ref{fig:measurement_scenario}).

The \gls{ctf} is extracted over a \gls{cpi} of 8 radio frames, consisting of 32 \gls{dmrs}-based estimates collected every 2.5\,ms.
Signal processing is performed within a real-time dApp \cite{Villa2026pp, DOro2022}, including the \gls{los}-based over-the-air synchronization to compensate the time, carrier and sampling frequency offsets as well as the phase offsets and gain mismatch between the \gls{gnb} array elements.
The remaining sensing processing follows the procedure described in Section~\ref{sec:procedure}.
For the target localization, we assume absolute \gls{los} delay knowledge determined from a known \gls{ue} location, as the \gls{5g} positioning function is not yet implemented in the testbed.

\begin{figure}[!t]
    \centering
    \subfloat[]{
        \centering
        \includegraphics[width=0.6\columnwidth]{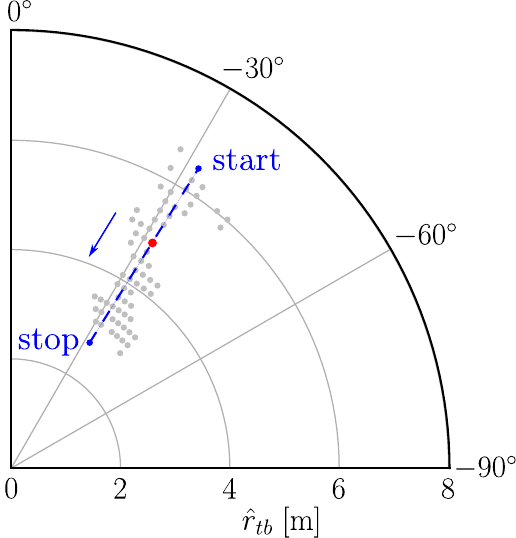}
        \label{fig:meas_rng_ang_map}
    }
    \\
    \subfloat[]{
        \centering
        \includegraphics[width=0.96\columnwidth]{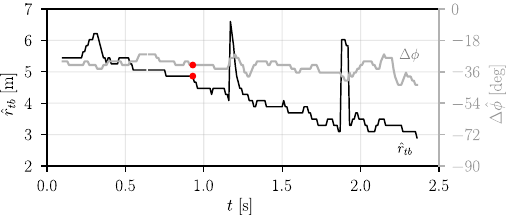}
        \label{fig:meas_rng_ang_time}
    }
    \caption{Experimental results showing a) detected target positions and b) corresponding distance and azimuth over time.}
    \label{fig:meas_rng_ang}
\end{figure}

The results are presented in  Fig.~\ref{fig:meas_rng_ang}, showing estimated target positions (Fig.~\ref{fig:meas_rng_ang_map}) and the corresponding target distance and azimuth angle variation over time (Fig.~\ref{fig:meas_rng_ang_time}) for the subject walking towards the \gls{ru}, with the dashed line indicating the walking path.
The detected target positions are clustered around the walking path, typically remaining within 0.5\,-\,1\,m distance from the actual subject position.
Note that the larger deviations, such as observed from the (monostatic) range curve in Fig.~\ref{fig:meas_rng_ang_time}, can be suppressed by employing appropriate tracking algorithms, not considered in this work.

\section{Conclusions}
\label{sec:conclusions}
This paper considers an opportunistic approach to enable bistatic sensing in evolved \gls{5g} \gls{fr1} networks, by exploiting uplink \gls{csi} estimates obtained from pilot signals.
The \gls{5g} \gls{ue} positioning function and \gls{gnb} antenna array are explored for \gls{los} absolute range and \gls{aoa} estimation, effectively allowing for target localization.

The presented sensing procedure is demonstrated based on simulation and measurements obtained with a real-time \gls{sdr}-based setup, developed using the NVIDIA Aerial open-source \gls{gnb} implementation and commercially-available \gls{ru} and \gls{ue}.
By considering an indoor scenario with a person walking, the results demonstrate the feasibility of target localization via uplink communication in \gls{5g} \gls{fr1} networks, with minimum required modifications.


\IEEEtriggeratref{12}


\end{document}

%% file: acronyms.tex
 
\newacronym{isac}{ISAC}{integrated sensing and communication}
\newacronym{3gpp}{3GPP}{3rd Generation Partnership Project}
\newacronym{5g}{5G}{fifth generation}
\newacronym{aoa}{AoA}{angle of arrival}
\newacronym{cfar}{CFAR}{constant false alarm rate}
\newacronym{cfo}{CFO}{carrier frequency offset}
\newacronym{cir}{CIR}{channel impulse response}
\newacronym{cpe}{CPE}{customer premises equipment}
\newacronym{cpi}{CPI}{coherent processing interval}
\newacronym{csi}{CSI}{channel state information}
\newacronym{ctf}{CTF}{channel transfer function}
\newacronym{dmrs}{DMRS}{demodulation reference signal}
\newacronym{fr1}{FR1}{frequency range 1}
\newacronym{gnb}{gNB}{next-generation NodeB}
\newacronym{dft}{DFT}{discrete Fourier transform}
\newacronym{idft}{IDFT}{inverse discrete Fourier transform}
\newacronym{los}{LoS}{line-of-sight}
\newacronym{music}{MUSIC}{multiple signal classification}
\newacronym{ofdm}{OFDM}{orthogonal frequency-division multiplexing}
\newacronym{prach}{PRACH}{Physical Random Access Channel}
\newacronym{ran}{RAN}{radio access network}
\newacronym{ru}{RU}{radio unit}
\newacronym{tx}{Tx}{transmitter}
\newacronym{rx}{Rx}{receiver}
\newacronym{sdr}{SDR}{software-defined radio}
\newacronym{sfo}{SFO}{sampling frequency offset}
\newacronym{snr}{SNR}{signal-to-noise ratio}
\newacronym{srs}{SRS}{sounding reference signal}
\newacronym{ta}{TA}{timing advance}
\newacronym{to}{TO}{timing offset}
\newacronym{ue}{UE}{user equipment}
\newacronym{ula}{ULA}{uniform linear array}
\newacronym{mac}{MAC}{medium access control}